\documentclass[letterpaper,12pt]{article}
\pdfoutput=1

\usepackage{jheppub}
\usepackage{amsmath}
\usepackage{graphicx}
\usepackage{subfig}
\usepackage{cancel}
\usepackage[dvipsnames]{xcolor}
\usepackage{color}
\usepackage{mathrsfs}
\usepackage{amsfonts}

\allowdisplaybreaks

\def\({\left(}
\def\){\right)}
\def\[{\left[}
\def\]{\right]}

\newcommand{\reef}[1]{(\ref{#1})}

\newcommand{\ban}[1]{\begin{align}#1\end{align}}

\newcommand{\gp}[1]{{g^{(2)}_{#1}}}
\newcommand{\mathbreak}{\notag\\}

\newcommand{\corr}[1]{\left< #1\right>}
\newcommand{\Tr}{\text{Tr}}

\arxivnumber{1605.02804}

\title{Holographic confinement in inhomogeneous backgrounds}

\author{Donald Marolf}
\author{and Jason Wien}

\affiliation{Department of Physics, University of California, Santa Barbara, CA 93106, USA}

\emailAdd{marolf@physics.ucsb.edu}
\emailAdd{jswien@physics.ucsb.edu}

\abstract{As noted by Witten, compactifying a $d$-dimensional holographic CFT on an $S^1$ gives a class of $(d-1)$-dimensional confining theories with gravity duals.  The prototypical bulk solution dual to the ground state is a double Wick rotation of the AdS$_{d+1}$ Schwarzschild black hole known as the AdS soliton.  We generalize such examples by allowing slow variations in the size of the $S^1$, and thus in the confinement scale. Coefficients governing the second order response of the system are computed
for $3 \le d \le 8$ using a derivative expansion closely related to the fluid-gravity correspondence.  The primary physical results are that i) gauge-theory flux tubes tend to align orthogonal to gradients and along the eigenvector of the Hessian with the lowest eigenvalue, ii) flux tubes aligned orthogonal to gradients are attracted to gradients for $d \le 6$ but repelled by gradients for $d \ge 7$, iii) flux tubes are repelled by regions where the second derivative along the tube is large and positive but are attracted to regions where the eigenvalues of the Hessian are large and positive in directions orthogonal to the tube, and iv) for $d > 3$,  inhomogeneities act to raise the total energy of the confining vacuum above its zeroth order value.}

\begin{document}
\maketitle

\section{Introduction}

The AdS/CFT correspondence \cite{maldholo, GKP, wittenholo} provides elegant geometrizations of many aspects of quantum field theory, including the phenomenon of confinement \cite{witten}. In gauge theories, an order parameter for confinement is the expectation value of a ``temporal'' Wilson loop around a Euclidean time circle:
\ban{
\corr{|\Tr(W)|} \equiv \corr{\frac 1 N \Tr\left( \mathcal P e^{- \oint A_\tau d\tau}\right)} \ \ \ . \label{Wloop}
}
The expectation is of the form $e^{-\beta E_q}$, where $E_q$ is the energy of a probe quark. In a confining phase this energy diverges, and so the expectation value \eqref{Wloop} vanishes. But a non-zero expectation value requires a finite probe quark energy and implies the theory to be in a deconfined phase \cite{polyloop,Tloop1,Tloop2}.

The holographic prescription  \cite{Wloop} for computing the expectation value of a Wilson loop $\mathcal C$ involves considering a fundamental string in the bulk which intersects the asymptotically AdS conformal boundary on the curve $\mathcal C$ defined by the Wilson loop.   Here we identify the (conformal) boundary of the bulk with the gauge theory spacetime.  At small bulk string coupling, the semi-classical approximation to
the associated worldsheet path integral gives
\ban{
\corr{|\Tr(W)|}  \simeq e^{-S_\text{cl}}, \label{hololoop}
}
where $S_{\text{cl}}$ is the classical string action of Euclidean worldsheet.  As we focus on the vanishing or non-vanishing of \reef{hololoop}, we need only determine if any worldsheets have finite action.   When there is no bulk worldsheet with boundary $\mathcal C$, expression \reef{hololoop} vanishes and the theory is confined.

The bulk topology associated with the Euclidean time circle is thus of critical importance.   When this circle is non-contractible, there can be no worldsheet with boundary $\mathcal C$ having the topology of a disk.  Since other topologies are allowed only in special cases\footnote{When the bulk has additional boundaries not associated with the original CFT spacetime.  Such boundaries typically lie at the end of an infinite throat related to an extreme horizon in the bulk}, it is of great interest to construct asymptotically locally AdS spacetimes with non-contractible  Euclidean time circles.

The prototypical example of a bulk geometry dual to a confining vacuum is the AdS-soliton \cite{witten,soliton}.  The solution may be constructed by Wick-rotating the Schwarzschild-AdS black hole and involves an arbitrary constant $b>0$.   In Fefferman-Graham gauge and Euclidean signature the metric may be written
\ban{
ds^2 =\frac{\ell^2} {z^2} \left[dz^2+ {\left(1+\frac {z^d}{b^d}\right)^{4/d}}  d\tau^2 + {\left(1+\frac {z^d}{b^d}\right)^{4/d}} dx_i dx^i+ \alpha_d^2 b^2 {\left(1-\frac{z^d}{{b}^d}\right)^2 \left(1+\frac{z^d}{{b}^d}\right)^{\frac{4}{d}-2}} d\theta^2 \right], \label{AdS-Soliton}
}
where $i=1, \dots d-2$ and $\alpha_d = \frac{2^{1-2/d}}{d}$.
We take $\theta$ to be dimensionless and to have $b$-independent period $2 \pi$ (as required by regularity at $z=b$). The conformal boundary may be taken to have metric
\ban{
\label{boundarymetric}
ds^2_{bndy} =   d\tau^2 +dx_i dx^i + \alpha_d^2 b^2\, d\theta^2,
}
so that $b$ controls the size of the $\theta$-circle on the boundary.

Below, we generalize this solution by allowing the size of the $S^1$ -- and thus the confinement scale -- to vary slowly along the boundary.  We work in Euclidean signature, but our results define Lorentz-signature solutions via a trivial Wick rotation of $\tau$, or equivalently by evolving the associated initial data at $t=0$.  We construct the bulk geometries in section \ref{method} using an adiabatic expansion.   Section \ref{physics} then extracts predictions for Wilson loops and the stress tensor in the dual gauge theory.  Readers most interested in such results may skip directly to this section. Numerical results for interesting coefficients are given for $3 \le d \le 8$.  The special case $d=2$ is solved analytically in appendix \ref{2D} and used to check our numerical codes.

\section{Adiabatically Varying Confining Vacua}
\label{method}

In any local theory, one may use a solution with continuous free parameters to build new solutions by promoting constant such parameters to slowly varying functions.  The explicit functional form will then require corrections, but these may be found by solving the equations of motion in an adiabatic expansion.  In particular, this procedure has been used extensively in the fluid-gravity correspondence \cite{FG} to construct holographic duals of conformal fluids near thermal equilibrium; see \cite{FGref1,FGref2} for reviews. Indeed, because \eqref{AdS-Soliton} is the double-Wick rotation of an AdS-Schwarzschild black hole, our solutions below could have been constructed as double Wick-rotations of appropriately static and symmetric instances of the fluid-gravity correspondence that satisfy certain regularity conditions.  However, we nevertheless find it useful to construct the relevant equations and study regularity directly in terms of coordinates adapted to our symmetries (as opposed to the ingoing Eddington-Finkelstein black hole coordinates of \cite{FG,FGref1,FGref2}).

To be more explicit, suppose that we begin with a bulk geometry having free parameters $\{c_\alpha\}$.   We promote each constant to a slowly varying function by making the replacement $c_\alpha \to c_\alpha(\epsilon x)$ to define a new metric $\tilde g^{(0)}_{AB}.$  Here $\epsilon$ is a dimensionless book-keeping parameter that controls the adiabatic expansion.

Our $\tilde g^{(0)}_{AB}$ no longer solves Einstein's equation exactly, but we can use it to construct a solution by considering the ansatz
\ban{\label{ansatz0}
ds^2 &=  \tilde g^{(0)}_{AB} \, dx^A dx^B +\epsilon\, \tilde g^{(1)}_{AB}\,  dx^A dx^B +\epsilon^2 \, \tilde g^{(2)}_{AB} \, dx^A dx^B +\cdots.
}
Inserting \eqref{ansatz0} into the Einstein equation gives, at each order $n$, a set of equations for the metric correction $\tilde g^{(n)}_{AB}$.
In general, at each order $n$ there may also be consistency conditions that impose relations between the $c_\alpha$ and their derivatives.  However, no such conditions will arise in the setting studied below.

We will use this method to construct a class of confining geometries  which approach the AdS-soliton \eqref{AdS-Soliton} in the limit as $\epsilon \to 0$.  Our solutions are constructed in Euclidean signature and have a $\tau$ translation symmetry.  As a result, they are bulk stationary points of the path integral that computes the vacuum of the dual gauge theory.  As in the discussion of \cite{witten,soliton} we assume this saddle to dominate.  Wick rotating to Lorentz signature or evolving initial data from $t=0$ will then give Lorentz-signature solutions dual to the gauge theory vacua.

\subsection{Ansatz and boundary conditions}

We begin with the AdS-soliton \eqref{AdS-Soliton} and promote $b$ to a slowly varying function of a single spatial coordinate $x$, i.e. $b \to b(\epsilon x)$. The effect on the boundary metric is to make the size of the $S^1$ fibers vary with $x$. Although for simplicity we will allow this size to vary only along a single coordinate direction, we describe at the end of section \ref{AdSol} below how at order $\epsilon^2$ this seemingly-special case in fact suffices to determine the response to completely general slow variations of $b$ in the $(d-1)$ directions $(\tau,x^i)$.

Since the dual CFT will have a ground state on any static spacetime, one expects no restrictions on the functional form of $b(\epsilon x)$. We will verify below that no constraints arise within the adiabatic expansion.   A key point will be that adding $x$-dependence in the above way will allow us to preserve regularity everywhere in the bulk, and in particular at the fixed points of the rotational Killing field $\partial_\theta$.

It will be convenient to let $x=x^1$, $y^1=  \tau$, and $y^i = x^i$ for $i \ge 2$.   With these definitions, the boundary coordinates are given by $x^\mu=(\theta, x, y^i)$ where again $i=1,\dots, d-2$.  Below, we use rotational invariance among the $y^i$ to write $g_{y^i y^j} = g_{yy} \delta_{ij}$.

Working in Fefferman-Graham gauge, we consider solutions of the form
\ban{
ds^2 &=\frac{\ell^2} {z^2} g_{AB} dx^A dx^B =  \frac {\ell^2} {z^2} \left( g^{(0)}_{AB} \, dx^A dx^B +\epsilon\, g^{(1)}_{AB}\,  dx^A dx^B +\epsilon^2 \, g^{(2)}_{AB} \, dx^A dx^B+\cdots \right) ,
\label{full}}
so that in the notation of \eqref{ansatz0} we have $\tilde g^{(n)}_{AB} = \frac{\ell^2}{z^2}g^{(n)}_{AB}$.  The explicit form of our zeroth order ansatz is
\ban{
g^{(0)}_{AB} \, dx^A dx^B =\,& dz^2 + \alpha_d^2 b^2 {\left(1-\frac{z^d}{{b}^d}\right)^2 \left(1+\frac{z^d}{{b}^d}\right)^{\frac{4}{d}-2}} d\theta^2\mathbreak
& + {\left( 1+\frac {z^d}{b^d}\right)^{4/d}} dx ^2 + {\left(1+\frac {z^d}{b^d}\right)^{4/d}} \sum_i dy^i dy^i.  \label{ansatz}
}

Using the Fefferman-Graham gauge condition $g^{(n)}_{Az}=0$ for $n\geq 1$ as well as reflection symmetry in both $\theta$ and $y^i$, shows that all $g^{(n)}_{AB}$ remain diagonal.  Similarly, only the $zz$, $zx$, $xx$, $yy$, and  $\theta \theta$  components of the Einstein tensor can be non-zero.

We wish to satisfy the vacuum Einstein's equation with a negative cosmological constant: \ban{
0=E_{AB} : = R_{AB} - \frac 12 R\, g_{AB}+\Lambda \, g_{AB} \ \ \ .
}
As in \cite{FG}, at each order in the adiabatic expansion we have
$\frac{d(d+1)}2$ equations $E^{(n)}_{\mu\nu}=0$ involving second derivatives with respect to $z$; we refer to these equations as dynamical.   Here $\mu,\nu$ range over all boundary coordinates.   We also obtain $d+1$ equations involving no more than first derivatives in $z$, and which we call constraints.  The latter divide themselves into
$E^{(n)}_{z\mu} =0$ and $E^{(n)}_{zz}=0$.   Rotational symmetry in the $y^i$ requires
$E^{(n)}_{y^i y^j}=E^{(n)}_{y y}\delta_{ij}$,
so at each order we have only three distinct dynamical equations $E^{(n)}_{xx}$, $E^{(n)}_{\theta\theta}$, and $E^{(n)}_{yy}$ for the three undetermined metric functions $g^{(n)}_{\theta\theta},g^{(n)}_{xx}$, and $g^{(n)}_{yy}$.  Moreover, each derivative $\partial_x$ adds another factor of $\epsilon$, so the dynamical equations for $g^{(n)}_{AB}$ are ultra-local in the boundary directions. We are left with three coupled second order ordinary differential equations in $z$.

The dynamical equations require two boundary conditions to fix the solution uniquely.  The first is given by fixing the induced metric on the boundary to be given by \reef{boundarymetric} with $b \to b(x)$. The zeroth order ansatz satisfies
\ban{
\label{bndymetric}
\lim_{z\to 0}  g^{(0)}_{\mu\nu} dx^\mu dx^\nu= dx^2 + \alpha_d^2 b^2 d\theta^2 + dy_i dy^i,
}and so gives the correct boundary metric to all orders.  We therefore impose
\ban{
\lim_{z\to 0} g^{(n)}_{\mu\nu} =0
}
for all $n >0$.

The second boundary condition is determined by regularity at the fixed points of $\partial_ \theta$.  This occurs at some $z=\tilde b(x)$ where the associated $S^1$ shrinks to zero size.  At zeroth order one finds $\tilde b=b$, though there are corrections at higher orders.  To impose regularity, it suffices to construct coordinates $R(z,x)$ and $X(z,x)$ such that $g_{\theta \theta}$ vanishes at $R=0$ and the metric takes the form
\ban{
ds^2=g_{RR}|_{R=0} \left(dR^2 + R^2 d\theta^2\right)+ g_{XX}|_{R=0} dX^2 + g_{YY}|_{R=0} \sum_{i=1}^{d-2} dY^i dY^i + O(R^2)\label{regular}
 }
where $g_{RR}|_{R=0}, g_{XX}|_{R=0}, g_{YY}|_{R=0}$ are positive (and thus non-vanishing) functions of $X$.

Expanding the zeroth-order ansatz \eqref{ansatz} in powers of $z-b(x)$ shows that it satisfies regularity as previously claimed.   One may then check that the full ansatz  \eqref{full} satisfies \eqref{regular} to order $\epsilon^2$ with
\ban{
z&=(1-R) b-\epsilon^2\, \frac 13 16^{-1/d}b \left( {b'}^2+\frac {2}{ \alpha_d^2d^2} \,  \left. \partial_z^2g_{\theta\theta}^{(2)}\right|_{z=b}\right)+O(\epsilon^4)\notag \\
x&=X+\epsilon \, 16^{-1/d}\, b \, b '  \left( R +\frac 1{2}  R^2-\frac 16(d-2) R^3   \right)+O(\epsilon^4, R^4) ,\label{regtrans}
}
so long as we impose the boundary conditions
\ban{
0&=\left. g_{\theta\theta}^{(1)} \right|_{z=b}\notag\\
0&=\left.{\partial_z g_{xx}^{(1)}}\right|_{z=b}\notag\\
0&=\left.{\partial_z g_{yy}^{(1)}}\right|_{z=b}\notag\\
0&= \left.{\partial_z g_{\theta\theta}^{(2)}}\right|_{z=b}-\frac{1}{6} b\left(\alpha_d^{2} d^2 {b'}^2+2 \left. {\partial_z^2 g_{\theta\theta}^{(2)}}\right|_{z=b}\right)\notag\\
0&= 2\, d \, b \left. g_{xx}^{(2)}\right|_{z=b} +2\alpha_d^{-2} \left. {\partial_z g_{\theta\theta}^{(2)}}\right|_{z=b}-d \, b^2\left. {\partial_z g_{xx}^{(2)}}\right|_{z=b}+ 2 \, d \,b^2\,  b''\notag\\
0&= 2 \,d  \, b\left. g_{yy}^{(2)}\right|_{z=b} + 2 \alpha_d^{-2} \left. \partial_z  {g_{\theta\theta}^{(2)}}\right|_{z=b}-d\, b^2 \left. \partial_z  {g_{yy}^{(2)}}\right|_{z=b}- 2 \, d \,b\, {b'}^2 .\label{bcs}
}
We emphasize that we have chosen the period of $\theta$ to remain precisely $2\pi$ at all $x$ at each order in $\epsilon$.

\subsection{Adiabatic solutions}
\label{AdSol}
We have now specified two boundary conditions at each order for each of the dynamical variables $g^{(n)}_{xx}, g^{(n)}_{\theta \theta},$ and $g^{(n)}_{yy}$.  This is enough to uniquely determine solutions to the dynamical equations $E_{\mu \nu}=0$ at each order.  It turns out that any such solution automatically satisfies the constraints $E_{z A}=0$ or, equivalently, $E_{RA} =0$.  For $A=\theta, Y^i$ this is clear from the reflection symmetries $\theta \rightarrow - \theta$ and $Y^i \rightarrow - Y^i$.  For $A=X, R$, we proceed by noting that  the Bianchi identities $\nabla_A E^{AB}$ imply first order evolution equations for the constraints $E^{RA}$.  Using \eqref{regular}, one finds that imposing $E^{\mu \nu}=0$ requires $E^{RR} = C^{RR}\left( R^{-1} + \dots \right)$ and $E^{RX} = C^{RX}\left( R^{-1} + \dots \right)$ where $C^{RR}, C^{RX}$ are constants and the dots ($\dots$) represent terms that vanish as $R\rightarrow 0$.  But regularity requires\footnote{A simple argument notes that ${\rm Tr} E^2 : = E^{AB}E^{CD} g_{DA}g_{BC}$ is a positive definite quadratic form that must be finite at $R=0$.  Explicitly, the leading terms at $R=0$ are $(g_{RR}E^{RR})^2 + 2 g_{RR} g_{XX} (E^{RX})^2$.  } $E_{RR}, E_{RX}$ to be finite at $R=0$.  This sets $C_{RR} =0 = C_{RX}$, so that the constraints hold identically everywhere in the bulk.  It thus suffices to solve the dynamical equations $E_{\mu \nu}=0$ alone subject to \eqref{bndymetric} and \eqref{bcs}.   At least in the adiabatic expansion, this verifies the expectation that bulk solutions exist for all profiles $b(\epsilon x)$.

Let us now examine in more detail the equations $E_{\mu \nu}^{(n)} =0$ that result from expanding $E_{\mu \nu}$ in powers of $\epsilon$ . In general, the lower order terms $g^{(n)}_{AB}$ in \eqref{full} lead to sources for the higher order terms.
As noted above, each boundary derivative contributes an explicit power of $\epsilon$.  Covariance requires each term in $E_{\mu \nu}$ to contain an even number of such derivatives,
so evaluating $E_{\mu \nu}$ on the zeroth-order ansatz \eqref{ansatz} alone can provide source terms only for $g^{(n)}_{AB}$ with $n$ even.

In particular, there can be no source terms at order $\epsilon$ so that the dynamical equations for
$g^{(1)}_{AB}$ are homogeneous.  Since the boundary conditions \eqref{bndymetric} and \eqref{bcs} are also homogeneous at this order, the unique solution is $g^{(1)}_{AB} =0$.

The story is more interesting at second order.  Explicit computation gives the following lengthy dynamical equations:
\ban{
0=&4 (d-2) z^{d+2} \left((d+1) b^d+(d-2) z^d\right){b'}^2 -4\, b (d-2) z^{d+2} \left(b^d+z^d\right) b''\mathbreak
&-4\, b^2 (d-4) z^{2 d}\gp{xx}-4 \,b^2 (d-4) (d-2) z^{2 d}\gp{yy}\notag\\
&+b^2 z \left(b^d+z^d\right) \left((d-7) z^d-(d-1) b^d\right) \partial_z \gp{xx}\mathbreak
&+b^2 (d-2) z \left(b^d+z^d\right) \left((d-7) z^d-(d-1) b^d\right)\partial_z \gp{yy}\notag\\
&+b^2 z^2 \left(b^d+z^d\right)^2 \partial_z^2 \gp{xx}+ b^2 (d-2) z^2 \left(b^d+z^d\right)^2 \partial_z^2 \gp{yy},\mathbreak
\mathbreak
0=&4\, (d-2) z^{d+2} \left(z^d-b^d\right)^3 \left(b^{2 d}+(d+1) b^d z^d+(d-2) z^{2 d}\right){b'}^2 \mathbreak
&-4 \alpha_d^{-2} \,z^{2 d} \left(b^d+z^d\right)^2 \left(-(2d^2-5d+4) b^{2 d}-2 (3 d-4) b^d z^d+ (d-4) z^{2 d}\right)\gp{\theta\theta}\mathbreak
&-4\, b^2 (d-2) z^{2 d} \left(z^d-b^d\right)^3 \left((d+4) b^d+(d-4) z^d\right) \gp{yy} \mathbreak
&-\alpha_d^{-2} z \left(b^d-z^d\right) \left(b^d+z^d\right)^3 \left((d-1) b^{2 d}-2 (3 d-4) b^d z^d+(d-7) z^{2 d}\right) \partial_z \gp{\theta\theta}\mathbreak
&+b^2 (d-2) z \left(b^d-z^d\right)^2 \left(z^{2 d}-b^{2 d}\right) \left((d-1) b^{2 d}+8 b^d z^d+(d-7) z^{2 d}\right)\partial_z \gp{yy}\mathbreak
&+\alpha_d^{-2} z^2 (b^d - z^d)^2 (b^d + z^d)^4 \partial_z^2 \gp{\theta\theta} +b^2 (d-2) z^2 \left(b^d-z^d\right)^4 \left(b^d+z^d\right)^2\partial_z^2 \gp{yy},\mathbreak
\mathbreak
0= &\,4 z^{d+2} \left(b^d-z^d\right)^2 \left(z^d-b^d\right) \left(d\, b^{2 d}+((d-2) d-6) b^d z^d+(d-3) (d-2) z^{2 d}\right) {b'}^2\mathbreak
&+2 \,b \,z^2 \left(b^d-z^d\right)^3 \left(b^d+z^d\right) \left(b^{2 d}+4 b^d z^d+(2 d-5) z^{2 d}\right) b''\mathbreak
&-4 \alpha_d^{-2}\,z^{2 d} \left(b^d+z^d\right)^2 \left(-(2d^2-5d+4) b^{2 d}-2 (3 d-4) b^d z^d+(d-4) z^{2 d}\right)\gp{\theta\theta}\mathbreak
&-4 \,b^2 z^{2 d} \left(z^d-b^d\right)^3 \left((d+4) b^d+(d-4) z^d\right) \gp{xx}\mathbreak
&-4\, b^2 (d-3) z^{2 d} \left(z^d-b^d\right)^3 \left((d+4) b^d+(d-4) z^d\right) \gp{yy}\mathbreak
&-\alpha_d^{-2}z \left(b^d-z^d\right) \left(b^d+z^d\right)^3 \left((d-1) b^{2 d}-2 (3 d-4) b^d z^d+(d-7) z^{2 d}\right)\partial_z \gp{\theta\theta}\mathbreak
&+b^2 z \left(b^d-z^d\right)^2 \left(z^{2 d}-b^{2 d}\right) \left((d-1) b^{2 d}+8 \,b^d z^d+(d-7) z^{2 d}\right) \partial_z \gp{xx}\mathbreak
&+b^2 (d-3) z \left(b^d-z^d\right)^2 \left(z^{2 d}-b^{2 d}\right) \left((d-1) b^{2 d}+8 \,b^d z^d+(d-7) z^{2 d}\right) \partial_z \gp{yy}\mathbreak
&+\alpha_d^{-2}z^2 \left(b^d-z^d\right)^2 \left(b^d+z^d\right)^4\partial_z^2 \gp{\theta\theta} + b^2 z^2 \left(b^d-z^d\right)^4 \left(b^d+z^d\right)^2\partial_z^2 \gp{xx}\mathbreak
& + b^2 (d-3) z^2 \left(b^d-z^d\right)^4 \left(b^d+z^d\right)^2\partial_z^2\gp{yy}.
\label{mess}
}
As a check, we can use \eqref{mess} to analytically compute the asymptotic expansion of $g^{(2)}_{xx}, g^{(2)}_{\theta \theta}, g^{(2)}_{yy}$ in powers of $z$.  Solving  \reef{mess} via the Frobenius method near $z=0$, for $d \ge 3$ we find
\ban{
\gp{\theta\theta}&= \alpha_d^2\frac{b\, b''}{d-1}z^2 + c_{\theta} z^d+O(z^{d+1}),\mathbreak
\gp{xx}&= \frac{b''}{b\,(d-1)}z^2 + c_{x} z^d+O(z^{d+1}),\mathbreak
\gp{yy}&= -\frac{b''}{b\,(d-1)(d-2)}z^2 + c_{y} z^d+O(z^{d+1}), \label{FG}
}
where the coefficients of $z^d$ are determined by the boundary conditions at the horizon.

On the other hand, for any boundary metric $\gamma_{\mu \nu}^{(0)}$, it is known (see e.g. \cite{Fischetti:2012rd}) that for $d \geq 3$ the $z^2$ coefficient in the expansion of $g_{\mu \nu}$ is given by
\ban{
\gamma^{(2)}_{\mu\nu} = -\frac {\ell^2}{d-2}\left(\mathcal R_{\mu\nu} -\frac1{2(d-1)} \mathcal R\gamma^{(0)}_{\mu\nu}\right), \label{FG2}
}
where $\mathcal R_{\mu\nu}$ is the Ricci tensor of $\gamma^{(0)}_{\mu\nu}$. Furthermore, the terms $z^n\gamma^{(n)}_{\mu\nu}$ with $3\le n<d$ involve higher numbers of derivatives and so vanish to order $\epsilon^2$ (and similarly for the $z^d \log z^2$ term for even $d> 2$; for $d=2$ the $z^2 \log z^2$ term vanishes identically).  As the boundary curvature is given by
\ban{
\mathcal R_{\theta\theta}= -\epsilon^2 \alpha_d^2\, b\, b'',  \ \ \
\mathcal R_{xx}= - \epsilon^2\, \frac{b''}b,  \ \ \
\mathcal R = -2\, \epsilon^2\,  \frac{b''}b, \label{boundcurv}
}
we see that \eqref{FG2} agrees with \eqref{FG}.

\begin{figure}[h!]
\centering
\subfloat[]{\includegraphics[width=0.45\textwidth]{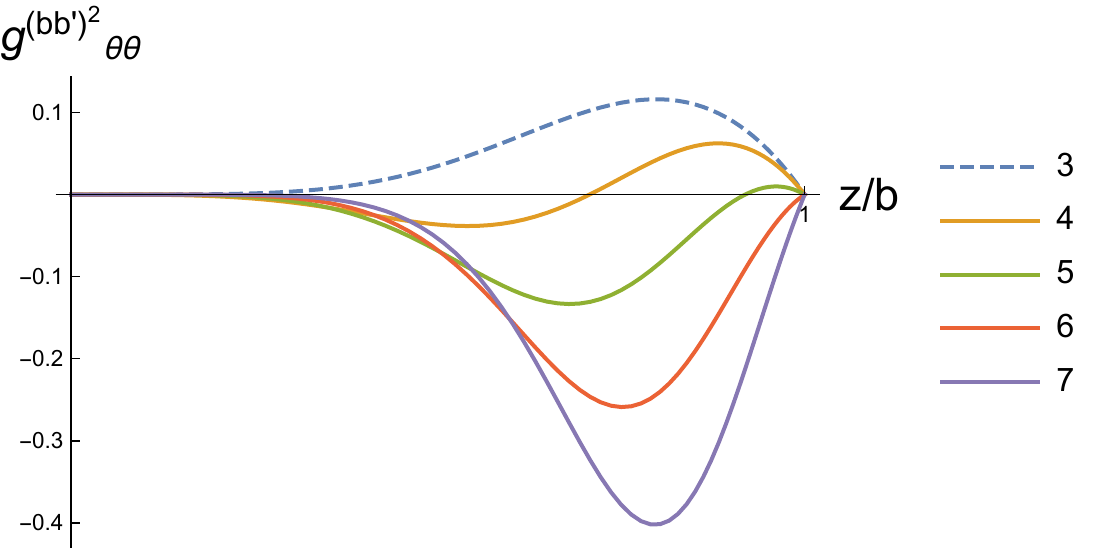}}\qquad
\subfloat[]{\includegraphics[width=0.45\textwidth]{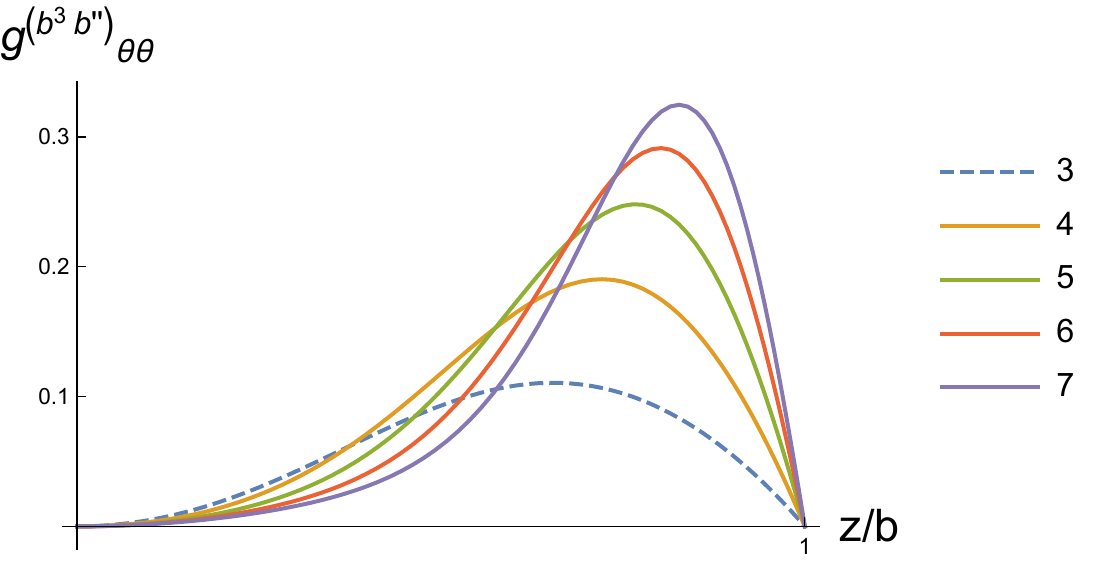}}
\caption{(Color online) Numerical solutions for (a) ${\sf g}_{\theta\theta}^{(bb')^2}$ and (b) ${\sf g}_{\theta\theta}^{(b^3b'')}$  as functions of $z/b$ for $d=3$ to $d=7$ using the notation \eqref{notation}. In each case the left endpoint is the asymptotic boundary $z=0$ and the right endpoint is the fixed point of $\partial_\theta$ (where $g_{\theta\theta}=0$).
}
\label{metricsoltt}
\end{figure}

\begin{figure}[h!]
\centering
\subfloat[]{\includegraphics[width=0.45\textwidth]{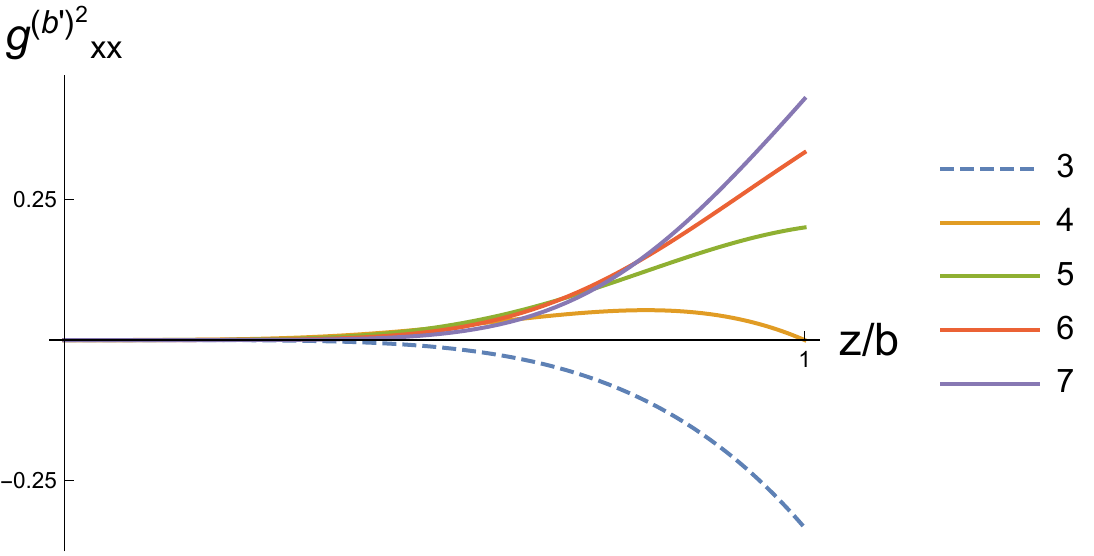}}\qquad
\subfloat[]{\includegraphics[width=0.45\textwidth]{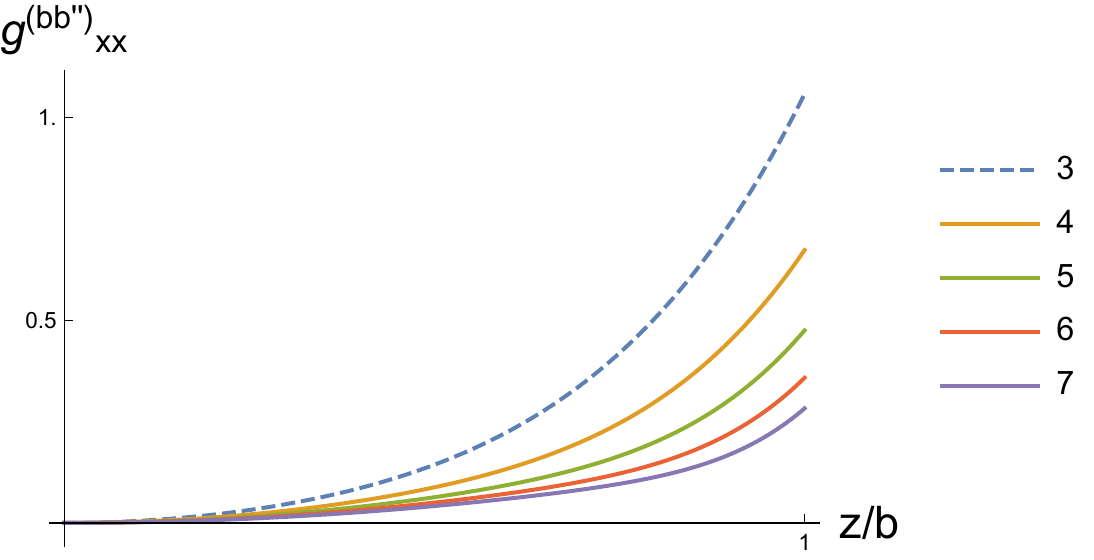}}
\caption{(Color online) Numerical solutions for (a) ${\sf g}_{xx}^{(b')^2}$ and (b) ${\sf g}_{xx}^{(bb'')}$  for $d=3$ to $d=7$ using the notation \eqref{notation}. In each case the left endpoint is the asymptotic boundary $z=0$ and the right endpoint is the fixed point of $\partial_\theta$ (where $g_{\theta\theta}=0$).
}
\label{metricsolxx}
\end{figure}

\begin{figure}[h!]
\centering
\subfloat[]{\includegraphics[width=0.45\textwidth]{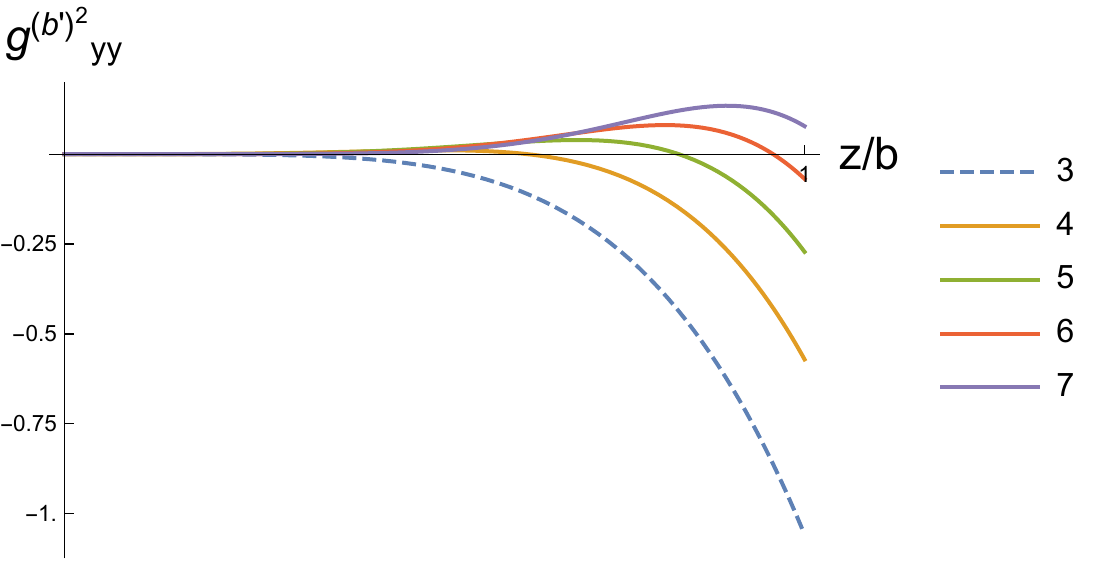}}\qquad
\subfloat[]{\includegraphics[width=0.45\textwidth]{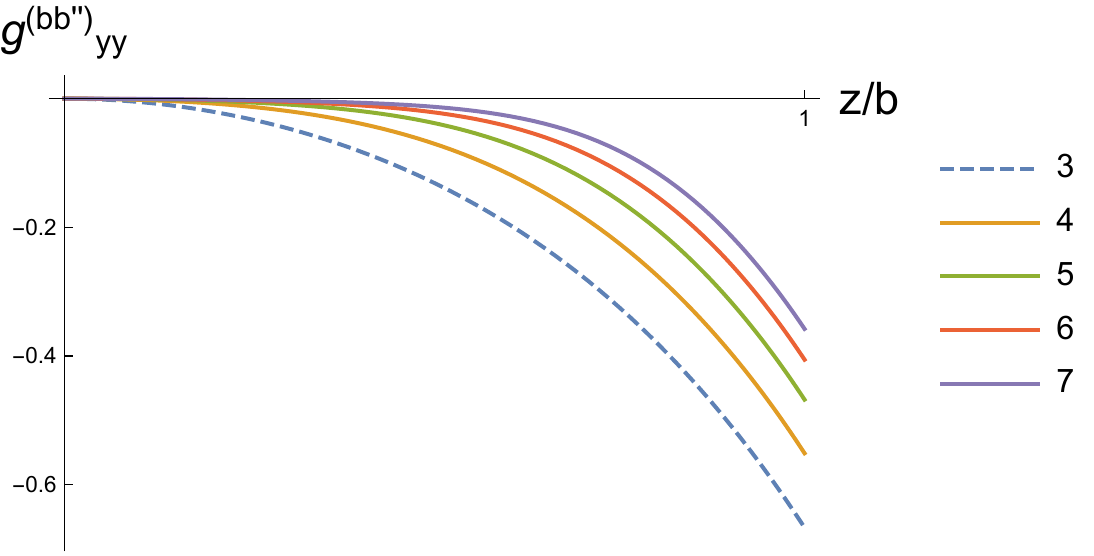}}
\caption{(Color online) Numerical solutions for (a) ${\sf g}_{yy}^{(b')^2}$ and (b) ${\sf g}_{yy}^{(bb'')}$  for $d=3$ to $d=7$ using the notation \eqref{notation}. In each case the left endpoint is the asymptotic boundary $z=0$ and the right endpoint is the fixed point of $\partial_\theta$ (where $g_{\theta\theta}=0$).
}
\label{metricsolyy}
\end{figure}

While the equations \eqref{mess} are highly coupled, they are also linear and can be solved numerically using the collocation methods described in \cite{numerics}.  By linearity, and dimensional analysis the solutions take the form
\begin{eqnarray}
\label{notation}
g^{(2)}_{xx}(z,x) &=&\left({b'(x)}\right)^2  {\sf g}_{xx}^{(b')^2}(z/b) +\left(b(x) {b''(x)}\right)
{\sf g}_{xx}^{(bb'')}(z/b)  , \cr
g^{(2)}_{yy}(z,x) &=& \left({b'(x)}\right)^2  {\sf g}_{yy}^{(b')^2}(z/b) +\left(b(x) {b''(x)}\right)
{\sf g}_{yy}^{(bb'')}(z/b) , \cr
g^{(2)}_{\theta\theta}(z,x) &=& \alpha_d^2 \left[ \left(b(x)\,b'(x)\right)^2 {\sf g}_{\theta\theta}^{(bb')^2}(z/b) +  b(x)^3\,b''(x) \
{\sf g}_{\theta\theta}^{(b^3b'')}(z/b) \right],
\end{eqnarray}
where the functions ${\sf g}_{xx}^{(b')^2}(z/b)$, etc  have no further dependence on $b(x)$.
  Results for these dimensionless coefficient functions are shown in figures \ref{metricsoltt} - \ref{metricsolyy}.

Although we have thus far allowed dependence only on a single coordinate $x$, the results above in fact determine the $O(\epsilon^2)$ response of our system to general slow variations of $b$ in the $(d-1)$ directions $(x,y^i)$.  In particular, since the metric at each order $\epsilon^n$ and each bulk point $(z, x, y^i, \theta)$ is locally determined by the boundary metric at $(x, y^i, \theta)$, in computing the response to gradients we are free to simply define $x$ at each such boundary point to run in the direction of any gradient of $b$, so long as we then take the $y^i$ to label the orthogonal directions.  We may then separately consider the response to the matrix of second derivatives of $b$ (the Hessian).  Here it is useful to choose coordinates that diagonalize the Hessian.  Furthermore, since the $O(\epsilon^2)$ response to second derivatives is linear, it suffices to separately compute the response to each eigenvalue $\lambda_\alpha$ of the Hessian.  And for studying any particular eigenvalue, we can choose the $x$ coordinate to run in the corresponding direction.  As a result, letting $\alpha,\beta$ run over directions corresponding to eigenvectors of the Hessian and denoting the the second order response  to the Hessian of $g_{AB}$ in the direction associated with some particular eigenvalue $\lambda_\beta$ by $g^{(2,Hess)}_{\beta \beta}$, we have
\begin{equation}
\label{Hessian}
g^{(2,Hess)}_{\beta \beta}
=b \left[ {\sf g}_{xx}^{(bb'')}(z/b) \lambda_\beta +
{\sf g}_{yy}^{(bb'')}(z/b) \sum_{\alpha \neq \beta} \lambda_\alpha \right]
\end{equation}
in terms of the functions ${\sf g}_{xx}^{(bb'')}(z/b)$, ${\sf g}_{yy}^{(bb'')}(z/b)$ computed above.

\section{Gauge Theory Implications}
\label{physics}

We now use the above solutions above to extract physical data about the confining gauge theory.
In particular, the quark/anti-quark potential $V(x_q,x_{aq})$ can be studied by computing the expectation value of rectangular Wilson loops extending along e.g. $x$ and $\tau=y^1$.  For $\Delta \tau\gg \Delta x = x_q-x_{aq}$, one expects from \eqref{Wloop} that
\ban{
W(\mathcal C) \sim e^{-V(x_q,x_{aq}) \, \Delta \tau}.
}
Using the holographic prescription \reef{hololoop}, we see that $V(x_q,x_{aq})$ is proportional to the (renormalized) area of the string world-sheet per unit time $\Delta \tau$. If we further take $\Delta x$ (and thus also $\Delta \tau$) much larger than the scale $b$, this renormalized area can be approximated by that of the corresponding rectangle on the hypersurface where $\partial_\theta =0$; we follow standard practice in referring to this surface as the IR floor. In the coordinate system \reef{regular}, the IR floor lies at $R=0$.  Transforming to Fefferman-Graham coordinates using \eqref{regtrans} and taking into account \eqref{bcs}, it also lies at $z=\tilde b$ with
\ban{
\tilde b &= b-\frac{\epsilon^2}{2}\, \, b^2\,  \alpha_d^{-2} \left. \partial_z g_{\theta\theta}^{(2)}\right|_{z=b}  +O(\epsilon^4).\label{floor}
}
Here we assume $d \ge 3$ so that there is at least one $y$ direction.  The special case $d=2$ is discussed separately in appendix \ref{2D}, where it is solved analytically and used to check our numerical codes.

We denote by ${\cal C}_{floor}$ the corresponding rectangular loop on this IR floor. Since $\tau=y^1$, the loop ${\cal C}_{floor}$  has area
\ban{
\label{xt}
A_{{\cal C}_{floor}} \approx \ell^2 \int dx\, dy^1 \, \left( \frac 1 {b^2} 16^{1/d} + \frac {\epsilon^2}{2 b^2} \left(h^{(2)}_{xx} +h^{(2)}_{yy}\right)\right),
}
where $\ell^2 h_{\mu \nu}$ is the induced metric on the IR floor. Similarly, for loops extending along $\tau$ and a $y$ direction, we have
\ban{
\label{yt}
A_{{\cal C}_{floor}} \approx \ell^2 \int dy^2\, dy^1 \, \left( \frac 1 {b^2} 16^{1/d} + \frac {\epsilon^2}{ b^2} \ h^{(2)}_{yy} \right).
}
The second order contributions to $h_{\mu\nu}$ are listed in the table in figure \ref{horizoncoeff} using notation analogous to \eqref{notation}.  Here we extend the calculations to $d=8$ due to an interesting change of sign for ${\sf h}^{({b'})^2}_{yy}$ between $d=6$ and $d=7$.
\begin{figure}[h!]
\centering
\begin{tabular}{c ||c |c || c | c }
$d$  & ${\sf h}^{({b'})^2}_{xx}$ & ${\sf h}^{({bb''})}_{xx}$ & ${\sf h}^{({b'})^2}_{yy}$& ${\sf h}^{({bb''})}_{yy}$\\
\hline
$3$ & -0.333 & 1.06 &-1.06& -0.667\\
$4$ & 0.00  & 0.673 &-0.571& -0.551\\
$5$ & 0.200  & 0.475 & -0.272 &-0.468\\
$6$ & 0.333  & 0.358 & -0.0688 & -0.406\\
$7$  & 0.429 & 0.282 & 0.0778 & -0.358\\
$8$  & 0.500 & 0.231 & 0.189 & -0.320
\end{tabular}

\caption{The coefficients ${\sf h}^{({b'})^2}_{xx}$, ${\sf h}^{({bb''})}_{xx}$, ${\sf h}^{({b'})^2}_{yy}$, and ${\sf h}^{({bb''})}_{yy}$
for the induced metric on the IR floor for various dimensions. Though we display only a few significant figures, estimating the numerical precision by comparing results for 100 and 150 lattice points suggests that our numerics are accurate to around a part in $10^{20}$. We note that ${\sf h}^{({b'})^2}_{xx}$ agrees with $(d-4)/d$ to the stated precision. }
\label{horizoncoeff}
\end{figure}

The factors in parentheses in \eqref{xt}, \eqref{yt} describe an effective tension for the gauge-theory flux tube whose stretching between the quark and anti-quark provides the confining potential.   Supposing for the moment that we allow $b$ to vary only in spatial directions ($x$ and $y^i$ for $i \ge 2$), the spacetime remains static and any flux tube will tend to orient itself to minimize this effective tension. As described at the end of section \ref{AdSol}, the coefficients above can be used to deduce the $O(\epsilon^2)$ response to general slow variations of $b$ across $(x,y^i)$.    The fact that ${\sf h}^{({b'})^2}_{xx} > {\sf h}^{({b'})^2}_{yy}$ for all $d$ in the table in figure \ref{horizoncoeff} thus implies that the flux tube tends to orient itself orthogonal to gradients.  In the same way, using \eqref{Hessian} and comparing directions associated with different eigenvalues of the Hessian, one sees that flux tubes also tend to align themselves with the lowest eigenvalue of the Hessian.

Interestingly, the change of sign of ${\sf h}^{({b'})^2}_{yy}$ between $d=6$ and $d=7$ means that a flux tube that succeeds in aligning itself orthogonal to gradients is attracted to strong gradients for $d \le 6$ but repelled from strong gradients for $d \ge 7$.  In all dimensions, flux tubes are repelled by regions where the second derivative along the tube would be large and positive but are attracted to regions where the eigenvalues of the Hessian are large and positive in orthogonal directions.

Another interesting piece of physics concerns the gravitational potential (or redshift) on the IR floor.  This is encoded in $h_{\tau\tau}=  h_{y^1y^1}  = (\frac1{b^2} {2^{4/d}} +\epsilon^2\, h^{(2)}_{yy}) + O(\epsilon^4)$. Again assuming a static spacetime one finds
\begin{equation}
h^{(2)}_{\tau\tau} = {\sf h}^{(b')^2}_{yy}   {|\partial_\mu b|^2}  + {\sf h}^{(bb'')}_{yy} {\rm Tr} \left(  b\,{\partial_\mu \partial_\nu b} \right),
\end{equation}
where $|\partial_\mu b|^2$  and ${\rm Tr} \left( \partial_\mu \partial_\nu b\right)$ respectively denote that norm of the gradient of $b$ and the trace of its Hessian. It is interesting that the table in figure \ref{horizoncoeff} shows gradients to lower the potential for $d \le 6$ but to raise the potential for $d=7,8$ (and presumably for higher dimensions as well).

Note that the value of $h_{\tau \tau}$ at an extremum (where $\partial_\mu b=0$) is unaffected by ${\sf h}^{(b')^2}_{yy}$. The fact that ${\sf h}^{(bb'')}_{yy} < 0$ in figure \ref{horizoncoeff} thus  means that the $O(\epsilon^2)$ corrections act to reduce the height of local maximum of $h_{\tau \tau}$ and to reduce the depth of local minima.  This should be no surprise, as at this order the response of the system is linear in $b''$ while on general grounds linear perturbation theory about the AdS soliton should describe the change in $h_{\tau \tau}$ as a smeared version of the boundary perturbation (i.e., given by convolution with some appropriate kernel) over a scale $\sim b$.  The point here is that smearing a maximum necessarily reduces its height, while smearing a minimum decreases its depth.  Indeed, all adiabatic coefficients associated with $b''$ can in principle be calculated from the associated linear-response Green's functions.

Finally, we can also compute coefficients for corrections to the boundary stress tensor.  Since at order $\epsilon^2$ we may neglect quadratic and higher powers of boundary curvatures, our boundary stress tensor takes the form
\ban{
T_{\mu\nu} &=  \frac {d\ell^{d-1}} {2\kappa}\gamma^{(d)}_{\mu\nu}  +O(\epsilon^4) \label{stresstensoreqn}
}
for both odd and even $d\ge 3$.  Here $\kappa =8\pi G_N / \ell^{d-1}$ in terms of the bulk Newton constant $G_N$ and $\gamma^{(n)}_{\mu\nu}$ is the $z^n$ coefficient of the Fefferman-Graham expansion (not to be confused with the $g^{(n)}_{\mu\nu}$ in the adiabatic expansion). We expand the stress tensor as
\ban{
T_{\mu\nu} & ={T_{\mu\nu} }^{(0)}+\epsilon \, {T_{\mu\nu}}^{(1)}+ \epsilon^2 \, {T_{\mu\nu} }^{(2)}+ \cdots
}
The zeroth order result is standard with
\ban{
T_{xx}^{(0)} &=\frac {\ell^{d-1}}{4 \pi G_N}  \frac{1}{b^{d}}\, ,\mathbreak
T_{yy}^{(0)} &=\frac {\ell^{d-1}}{4 \pi G_N}  \frac{1}{b^{d}}\, ,\mathbreak
T_{\theta\theta}^{(0)} &= -\frac {\ell^{d-1}}{4 \pi G_N}  \frac{\alpha_d^2 (d-1)}{b^{d-2}}\, .
}
Since $g^{(1)}_{\mu\nu}$ vanishes, so does ${T_{\mu\nu} }^{(1)}$.
 The second order contributions can be extracted from the numerical solutions for $g^{(2)}_{\mu\nu}$.  The results are summarized in figure \ref{stresstensor} using the notation
 \begin{eqnarray}
\label{Tnotation}
T^{(2)}_{xx} &=& \frac{\ell^{d-1}}{8 \pi G}\left[\left( \frac{ \left(b'\right)^2}{b^d}\right)  {\sf T}_{xx}^{(b^{-d} b'{}^2)} +
\left(\frac{b''}{b^{d-1}}\right)
{\sf T}_{xx}^{(b^{-(d-1)}b'')}  \right], \cr
T^{(2)}_{yy} &=& \frac{\ell^{d-1}}{8 \pi G}\left[\left( \frac{ \left(b'\right)^2}{b^d}\right)  {\sf T}_{yy}^{(b^{-d} b'{}^2)} +
\left(\frac{b''}{b^{d-1}}\right)
{\sf T}_{yy}^{(b^{-(d-1)}b'')}  \right], \cr
T^{(2)}_{\theta \theta} &=& \frac{\ell^{d-1}}{8 \pi G}\left[\left( \frac{ \left(b'\right)^2}{b^{d-2}}\right)  {\sf T}_{\theta\theta}^{(b^{-(d-2)} b'{}^2)} +
\left(\frac{b''}{b^{d-3}}\right)
{\sf T}_{\theta\theta}^{(b^{-(d-3)}b'')}  \right].
\end{eqnarray}

\begin{figure}[h!]
\centering
\begin{tabular}{c ||c |c || c | c || c | c}
$d$  & ${\sf T}_{\theta\theta}^{(b^{-(d-2)} b'{}^2)}$ & ${\sf T}_{\theta \theta}^{(b^{-(d-3)} b'')}$ & ${\sf T}_{xx}^{(b^{-d} b'{}^2)}$ & ${\sf T}_{xx}^{(b^{-(d-1)} b'')}$ & ${\sf T}_{yy}^{(b^{-d} b'{}^2)}$ & ${\sf T}_{yy}^{(b^{-(d-1)} b'')}$ \\
\hline
$3$ & 0.00 & 0.00 & 0.00 & 0.00 & 0.00 & 0.00\\
$4$ & $-0.375$ & 0.250 & 1.00  & 0.00 &1.00&$ -1.00$\\
$5$ & $-0.844$ & 0.422 & 2.30  & 0.00 & 2.30 &$-1.53$\\
$6$ & $-1.32$ & 0.529 & 3.78  & 0.00 & 3.78 & $-1.89$\\
$7$ & $-1.77 $ & $0.591$ & 5.38 & 0.00 & 5.38 & $-2.15$\\
$8$ & $-2.19 $ & $0.625$ & 7.07 & 0.00 & 7.07 & $-2.36$
\end{tabular}

\caption{The coefficients of the second order contributions to the boundary stress tensor for $3 \le d \le 8$. Estimating the numerical precision by comparing results for 100 and 150 lattice points suggests that our numerics are accurate to around a part in $10^{8}$.   To this accuracy our results satisfy $T_{yy}^{(b^{-d} b'{}^2)} =-\frac{d-2}{2} T_{yy}^{(b^{-(d-1)} b'')}$ and $T_{xx}^{(b^{-d} b'{}^2)}=T_{yy}^{(b^{-d} b'{}^2)}$.}
\label{stresstensor}
\end{figure}

As in our discussion of the potential on the IR floor, the signs of ${\sf T}_{\theta \theta}^{(b^{-(d-3)} b'')}$ and ${\sf T}_{yy}^{(b^{-(d-1)} b'')}$  are in all cases consistent with the idea that linear response tends to simply average over a scale of order $b$.  As a result, the $O(\epsilon^2)$ correction to the (negative) energy density of the confining vacuum makes this energy less negative at a local minimum of $b$ but more negative at a local maximum. On the other hand, gradients always make this energy density even more negative when the second derivatives are held fixed.

Of particular interest is the $O(\epsilon^2)$ shift $E^{(2)}$ in the total energy of the vacuum.  This is given by integrating $-T^{(2)}_{yy}$ over the boundary at $\tau =0$. The interesting point here is that first and second derivatives are often related when averaged over this surface. Indeed, imposing either a boundary condition $b \rightarrow constant$ as $x \rightarrow \pm \infty$ or periodic boundary conditions in $x$, integrating by parts gives

\begin{eqnarray}
\label{E2}
E^{(2)} &=& - \int_{bndy @ \tau=0} \sqrt{\sigma} \,T_{yy} = -  \frac{2\pi \ell^{d-1}}{8 \pi G} \int dx d^{d-2} y \ \alpha_d b \ \left[\left( \frac{ \left(b'\right)^2}{b^d}\right)  {\sf T}_{yy}^{(b^{-d} b'{}^2)} +
\left(\frac{b''}{b^{d-1}}\right)
{\sf T}_{yy}^{(b^{-(d-1)}b'')}  \right]  \cr
&=& -  \frac{\alpha_d \ell^{d-1}}{4 G} \int dx d^{d-2} y \left( \frac{ \left(b'\right)^2}{b^{d-1}}\right) \left[ {\sf T}_{yy}^{(b^{-d} b'{}^2)} + (d-2)
{\sf T}_{yy}^{(b^{-(d-1)}b'')}  \right],
\end{eqnarray}
where $\sqrt{\sigma} = \alpha_d b$ is the volume element on the $\tau=0$ slice of the boundary. As shown in figure \ref{vacenergy2}, the factor in square brackets is negative in all cases.  So the net effect of spatial variations is in fact to make $E^{(2)}$ positive, shifting the energy of the confined vacuum toward zero from its negative zeroth-order value.
\begin{figure}[h!]
\centering
\begin{tabular}{c ||c }
$d$  &  $\sf E^{(2)}$ \\
\hline
$3$ & 0.00 \\
$4$ & $-1.00$\\
$5$ & $-2.30$ \\
$6$ & $-3.78$  \\
$7$ & $-5.38 $  \\
$8$ & $-7.07 $
\end{tabular}
\caption{The coefficient ${\sf E^{(2)}}={\sf T}_{yy}^{(b^{-d} b'{}^2)}+(d-2){\sf T}_{yy}^{(b^{-(d-1)} b'')}$ of the second order contribution to the vacuum energy for $3 \le d \le 8$. The numerical precision is as in figure \ref{stresstensor}.}
\label{vacenergy2}
\end{figure}

It would be interesting to perform a similar analysis of the deconfined state.  Computing the second order shift in its free energy and comparing with \eqref{E2} would then determine whether the net effect of gradients is to increase  the deconfinement temperature $T_D$ at  $O(\epsilon^2)$, or to decrease $T_D$ as our results would appear to suggest.  Other interesting extensions would be to add additional curvature on the boundary.  Note that the particularly simple class of boundary metrics of the form
\begin{equation}
\label{kmet}
ds^2_{bndy} =   dx^2 + k^2(\epsilon x) dy_i dy^i + \alpha_d^2 b^2(\epsilon x)\, d\theta^2,
\end{equation}
is related to those studied here by a combination of a conformal transformation and a change of coordinates in the $x$ direction (associated with $dx \rightarrow dx/k$), so that the adiabatic coefficients associated with \eqref{kmet} can be computed analytically from the results given above.

\section*{Acknowledgements}
It is a pleasure to thank Eric Mefford, Sebastian Fischetti, and William Kelly for useful discussions.  We are especially grateful to Jorge Santos for conversations that originally led to this project and for his encouragement to focus on physical implications. This work was supported in part by the U.S. National Science Foundation under grant numbers PHY12-05500 and PHY15-04541 and by funds from the University of California.


\appendix

\section{2+1 Dimensional Bulk}\label{2D}

Due to the lack of local gravitational degrees of freedom in 2+1 dimensions, all complete asymptotically locally AdS spacetimes are diffeomorphic to global AdS$_3$ (or to a quotient thereof).  We can use this fact to analytically perform the $d=2$ analogue of the construction in section \ref{method}, which we can then use to check our numerical code.  The $d=2$ version of the Euclidean metric \reef{ansatz} is obtained by simply deleting the $y^i$ terms:
\ban{
ds^2=\frac{\ell^2} {z^2} \left[dz^2+\frac{b^2}{4} {\left(1-\frac{z^2}{{b}^2}\right)^2 d\theta^2 + \left(1+\frac {z^2}{b^2}\right)^{2}} dx^2\right]. \label{BTZ}
}

The adiabatic expansion proceeds just as in section \ref{method}.  We need only set $d=2$ in \eqref{mess} to find the dynamical equations
\ban{
0&=\left(b^2+z^2\right) \left(z \left(b^2+z^2\right) \partial_z^2 \gp{xx}-\left(z^2-3 b^2\right)\partial_z\gp{xx}\right)-8 b^2 z\, \gp{xx}, \mathbreak
0&=\left(b^2-z^2\right) \left(z \left(b^2-z^2\right)\partial_z^2 \gp{\theta\theta}+\left(z^2+3 b^2\right) \partial_z \gp{\theta\theta}\right)+8 b^2 z\ \gp{\theta\theta} . \label{2Ddiff}
}
We again have the boundary conditions
\ban{
\lim_{z\to 0} \, z^2 g^{(n)}_{\mu\nu}=0
}
at the asymptotic boundary, and regularity at fixed points of $\partial_\theta$ requires
\ban{
0&= \left.{\partial_z g_{\theta\theta}^{(2)}}\right|_{z=b}-\frac{1}{3} b\left( \frac{{b'}^2}{2}+ \left. {\partial_z^2 g_{\theta\theta}^{(2)}}\right|_{z=b}\right)\ \ \ ,\notag\\
0&= 2  \, b \left. g_{xx}^{(2)}\right|_{z=b} +4 \left.{\partial_z g_{\theta\theta}^{(2)}}\right|_{z=b} - \, b^2\left. {\partial_z g_{xx}^{(2)}}\right|_{z=b} + 2 \, b^2\, b'' \ \ \ .
\label{2bcs}
}
Solving \reef{2Ddiff}, \eqref{2bcs} yields
\ban{\label{2res}
\gp{\theta\theta}&= \frac{z^2 \left(b^2-z^2\right) {b'}^2}{8\, b^2} \ \ \ , \mathbreak
\gp{xx}&=\frac{z^2 \left(b^2+z^2\right) \left(2 \, b \,b''-{b'}^2\right)}{2 b^4} \ \ \ .
}
Setting $d=2$ in our numerical code gives solutions to \reef{2Ddiff}, \eqref{2bcs}  that agree with \eqref{2res} to one part in $10^{21}$.



\bibliographystyle{jhep}
	\cleardoublepage
\phantomsection
\renewcommand*{\bibname}{References}

\bibliography{references}

\end{document}